\documentclass[aps,prb,twocolumn,superscriptaddress,longbibliography]{revtex4-2}

\usepackage[utf8]{inputenc}
\usepackage{physics}
\usepackage{comment}
\usepackage{amssymb}
\usepackage{multirow}
\usepackage[colorlinks=true,linkcolor=blue,citecolor=blue,urlcolor=blue]{hyperref} 
\usepackage{graphicx}
\usepackage{amsmath}
\usepackage{dcolumn}
\usepackage{bm}
\usepackage{graphicx}
\usepackage{amsmath}
\usepackage{amsfonts}
\usepackage{color}
\usepackage{mathtools}
\usepackage{braket}
\usepackage{gensymb}
\usepackage{textcomp}
\usepackage{soul}

\begin{document}

\title{Interplay of altermagnetism and pressure in hexagonal and orthorhombic MnTe}

\author{Nayana Devaraj}
\email{nayanad@iisc.ac.in}
\affiliation{Solid State and Structural Chemistry Unit, Indian Institute of Science, Bangalore 560012, India}
\author{Anumita Bose}
\affiliation{Solid State and Structural Chemistry Unit, Indian Institute of Science, Bangalore 560012, India}
\author{Awadhesh Narayan}
\email{awadhesh@iisc.ac.in}
\affiliation{Solid State and Structural Chemistry Unit, Indian Institute of Science, Bangalore 560012, India}

\date{\today}



\begin{abstract}
Alternative magnetic materials or ``altermagnets", characterized by their \textcolor{black}{unconventional} non-relativistic, momentum-dependent spin-split states, represent a cutting-edge advancement in the field of \textcolor{black}{collinear} magnetism, offering promising avenues for spintronic applications. Among these materials, hexagonal MnTe has emerged as a standout material candidate for its substantial spin-splitting. In this study, employing first-principles electronic structure calculations and spin group symmetry analysis, we delve into the interplay of altermagnetism and pressure in two main phases of MnTe. Our relativistic calculations demonstrate the presence of tunable anomalous Hall effect (AHE) in hexagonal MnTe. In addition, our results underscore the pivotal role of pressure as a tuning parameter for the alternative magnetic traits in the system. Furthermore, we identify another phase of MnTe with orthorhombic structure, namely $\gamma$-MnTe, hosting altermagnetic characteristics. We study, in detail, its response in AHE and spin-spliting due to magnetization and pressure variations, respectively. Our study highlights the substantial impact of pressure on the properties of alternative magnetic materials, particularly emphasizing the pronounced tuning effect observed in the hexagonal and orthorhombic MnTe.
\end{abstract} 
\maketitle

\newpage

\section{Introduction}

In past few years significant progress has emerged in understanding the breaking of time-reversal symmetry~\cite{vsmejkal2020crystal,mazin2021prediction,vsmejkal2022anomalous} and spin-splitting within electronic bands~\cite{vsmejkal2020crystal,mazin2021prediction,ahn2019antiferromagnetism,naka2019spin,hayami2019momentum,yuan2020giant,hayami2020bottom,yuan2021prediction,ma2021multifunctional} in materials previously labeled as antiferromagnets due to their collinear-compensated magnetic order. These characteristics, typically associated with ferromagnetic materials, challenged conventional classifications \textcolor{black}{for collinear magnetic materials}. Subsequent experimental observations validated several of these predictions~\cite{feng2022anomalous, bose2022tilted}. \v{S}mejkal and co-workers addressed these ambiguities by employing a symmetry approach based on a nonrelativistic spin-group formalism~\cite{vsmejkal2022beyond,vsmejkal2022emerging,betancourt2023spontaneous,brinkman1966theory,litvin1974spin}. Through the utilization and refinement of the spin-group formalism, they elucidated three distinct phases of nonrelativistic collinear magnetism~\cite{vsmejkal2022beyond}. In the first phase, there exists either a singular spin lattice or opposite-spin sublattices lacking connections through any symmetry transformation, representing conventional ferromagnetism (or ferrimagnetism)~\cite{landau1946electrodynamics}. In contrast, the second phase is defined by opposite-spin sublattices interconnected via translation or inversion symmetries, indicative of conventional antiferromagnetism~\cite{brinkman1966theory,neel1971magnetism,corticelli2022spin}. In the newly-discovered third phase, opposite-spin sublattices are connected solely through rotations, without connections via translation or inversion. The third phase differs from both the conventional ferromagnetic phase, which exhibits nonrelativistic magnetization and spin-split bands that break time-reversal symmetry, and the conventional antiferromagnetic phase, which features nonrelativistic spin-degenerate bands that are invariant under time-reversal and has zero net magnetization. In this third phase, the spin-up and spin-down energy isosurfaces in the band structure are split but equally populated, effectively breaking time-reversal symmetry while maintaining a vanishing net magnetization -- this is referred to as ``altermagnetism", indicating the presence of symmetry-breaking spin-splitting with alternating signs in both real-space crystal and momentum-space electronic structures~\cite{vsmejkal2022beyond,vsmejkal2022emerging}. 

The emergence of altermagnetism has resulted in the reclassification of previously known materials, such as RuO$_2$, MnF$_2$, CrSb, and Mn$_5$Si$_3$, which were traditionally categorized as antiferromagnets~\cite{vsmejkal2022beyond,vsmejkal2022emerging,fedchenko2024observation,tschirner2023saturation,bhowal2024ferroically,reimers2024direct,leiviska2024anisotropy,badura2024observation}. Altermagnetism has been predicted in a large number of materials, both two-dimensional (2D) and three-dimensional (3D) crystals, with a wide variety of properties, ranging from insulators, semiconductors, and semi-metals to metals and superconductors~\cite{vsmejkal2022beyond,vsmejkal2022emerging,guo2023spin}. The structural and chemical diversity of altermagnets is extensive, and their unique properties, including zero net magnetization, make them highly stable and insensitive to external magnetic fields. The unique combination of zero net magnetization, spin-dependent properties, and structural diversity in altermagnets renders them potentially well-suited for energy-efficient information storage and processing in spintronic devices~\cite{vsmejkal2022beyond,vsmejkal2022emerging,khalili2024prospects,bai2024altermagnetism}. Several interesting effects are predicted in altermagnets, such as the anomalous Hall effect~\cite{vsmejkal2020crystal,vsmejkal2022anomalous}, crystal magneto-optical Kerr effect~\cite{mazin2021prediction,samanta2020crystal}, large nonrelativistic spin-splitting~\cite{vsmejkal2020crystal,yuan2020giant,osumi2024observation}, spin-polarized longitudinal and transverse currents~\cite{naka2019spin, gonzalez2021efficient}, and giant and tunneling magnetoresistance~\cite{vsmejkal2022giant,shao2021spin}. There is rapid ongoing activity to identify new altermagnetic materials, explore and tune their properties, and incorporate them into various applications~\cite{occhialini2022strain,liu2023inverse,bai2023efficient,kim2024observation,tschirner2023saturation,chakraborty2024strain}.

The anomalous Hall effect (AHE) has been a fascinating topic in condensed matter and materials physics since its discovery. It involves generation of a non-dissipative current, perpendicular to the applied electric field and Hall vector in the absence of an external magnetic field~\cite{nagaosa2010anomalous,vsmejkal2022anomalous}. The non-dissipative Hall current requires the breaking of time-reversal symmetry~\cite{vsmejkal2022anomalous}. Since the discovery of the AHE, the majority of studies have concentrated on ferromagnets, where time-reversal symmetry, $\tau$, is broken intrinsically~\cite{park2022anomalous}. Subsequently, the effect was also found in non-collinear antiferromagnets~\cite{chen2014anomalous,kiyohara2016giant,nayak2016large}. However, in \textcolor{black}{conventional} collinear antiferromagnets, the AHE is forbidden due to the presence of $\tau$ or $P\tau$ symmetry ($P$ is parity or inversion)~\cite{jungwirth2016antiferromagnetic}. \textcolor{black}{In contrast, despite having collinear antiparallel magnetic orientations, altermagnetic materials exhibit a different scenario compared to conventional collinear antiferromagnets regarding the aforementioned symmetry breaking, primarily due to the arrangement of adjacent non-magnetic atoms. This enables altermagnets behave like an unconventional antiferrromagnet featuring non-zero AHE~\cite{betancourt2023spontaneous,vsmejkal2022emerging}}. AHE has been observed and predicted in many altermagnetic materials, including RuO$_2$~\cite{vsmejkal2020crystal,tschirner2023saturation}, $\kappa$-(BETD-TTF)$_2$Cu[N(CN)$_2$]Cl~\cite{naka2020anomalous} and Mn$_5$Si$_3$~\cite{leiviska2024anisotropy}. The realization of AHE depends crucially on the magnetization vector orientation~\cite{betancourt2023spontaneous}, and the exploration of AHE altermagnetic materials remains a topic of active research.

Hexagonal MnTe, known as $\alpha$-MnTe, is a notable altermagnetic material~\cite{vsmejkal2022beyond}, recognized for its unique electrical and magnetic properties even before its classification as an altermagnet. The incomplete $d$-shell of Mn, coupled with its semiconducting nature and high Neel temperature of nearly 300 K, make $\alpha$-MnTe a particularly intriguing material~\cite{mimasaka1987pressure}. The detection of a spontaneous anomalous Hall signal in an epitaxial film of $\alpha$-MnTe, in the absence of an external magnetic field, highlights the time-reversal breaking phenomenon in this material~\cite{betancourt2023spontaneous}. Furthermore, strongly anisotropic band-splitting has been observed in both first-principles calculations and angle-resolved photoemission spectroscopy (ARPES) measurements~\cite{osumi2024observation,hajlaoui2024temperature}. The large band-splitting in $\alpha$-MnTe makes it promising for potential spintronic applications~\cite{rooj2024hexagonal, osumi2024observation}.

\begin{figure*}[t]
\centerline{\includegraphics[scale=.4]{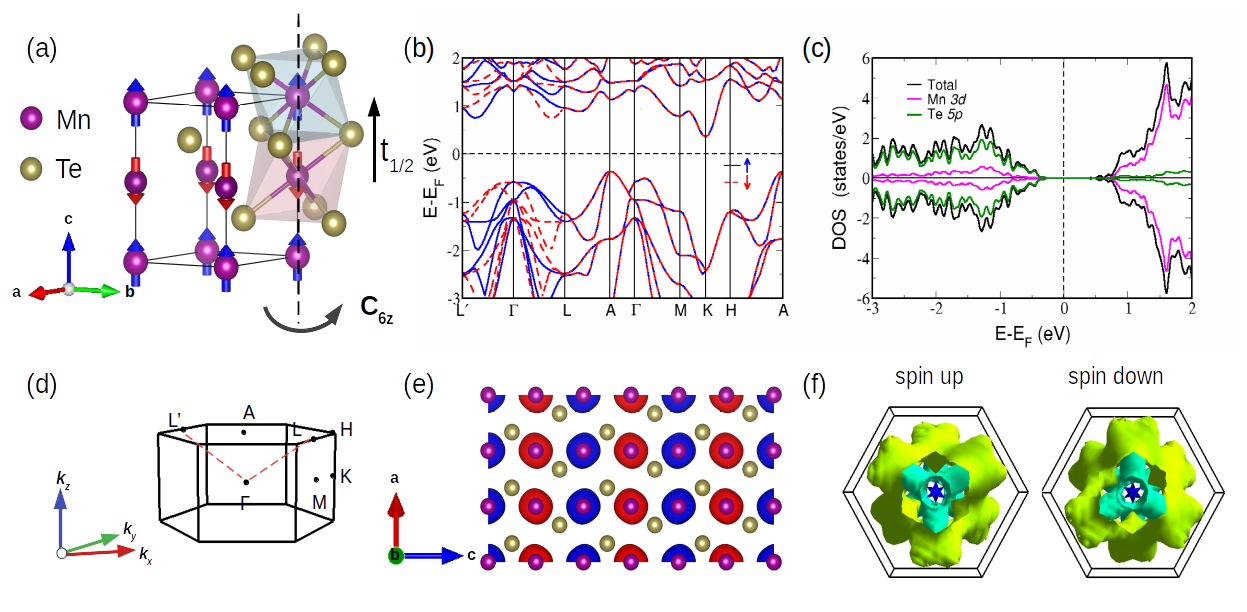}}
\caption{\textbf{Altermagnetism in $\alpha$-MnTe.} (a) Crystal structure of $\alpha$-MnTe, with the up and down spin sublattices distinguished by blue and red shading, respectively. The sublattices are connected by a non-symmorphic six-fold screw-axis rotation. (b) The spin-polarized electronic band structure along the high-symmetry path $L'-\Gamma-L-A-\Gamma-M-K-H-A$ is presented. Spin-up and spin-down bands are represented by blue solid lines and red dashed lines, respectively. Spin-splitting is evident along the path $L'-\Gamma-L$, while the bands for the two spin channels remain degenerate along $A-\Gamma-M-K-H-A$. (c) The spin-polarized density of states. Total density of states are depicted by black lines, while contributions from Mn 3$d$ orbitals and Te 5$p$ orbitals are represented by magenta and green lines, respectively. Mn $d$ orbitals exhibit a predominant contribution in the conduction manifold, whereas Te $p$ orbitals dominate in the valence manifold. (d) The bulk hexagonal Brillouin zone of $\alpha$-MnTe is illustrated, with high-symmetry points denoted by black dots. The $k$-path where spin-splitting is seen is marked by red dashed lines. (e) The magnetization density distribution for $\alpha$-MnTe is shown, with opposite-spin sublattices connected by a six-fold rotation about the $c$-axis. (f) Depicts the constant energy surfaces for up and down spins, connected via a six-fold rotation. Different colors represent different bands.}
\label{fig_MnTe_hex_plots}
\end{figure*}

In this study, through the state-of-the-art density functional calculations, we explore the interplay of altermagnetism and pressure in MnTe. We discover a remarkable enhancement of spin-splitting in hexagonal MnTe doubling from nearly 1 eV to 2 eV under applied pressure. We carry out a comprehensive investigation of the dependence of magnetic orientation of the relativistic band structure and AHE by combining spin group analysis with density functional calculations. We demonstrate that the anomalous Hall conductivity (AHC) can exceed 370 S/cm in $\alpha$-MnTe. Although MnTe adopts the hexagonal NiAs-type structure under ambient conditions, previous studies have revealed its phase transition to the orthorhombic phase under high pressure~\cite{mimasaka1987pressure,wang2022concurrent}, similar to other NiAs-type compounds~\cite{eto2001pressure,menyuk1969effects,onodera1999structural}. The orthorhombic phase of MnTe, known as $\gamma$-MnTe, holds significance in this context. Despite its importance, the electronic and magnetic properties of the orthorhombic phase of MnTe remain unexplored. Therefore, we expand our investigation to encompass the electronic and magnetic properties of different magnetic phases of the $\gamma$-MnTe. Our study reveals that one of the phases of $\gamma$-MnTe exhibits altermagnetism, thus expanding the library of altermagnetic materials. Furthermore, we predict the presence of a large and tunable AHE in $\gamma$-MnTe, where the direction of the AHC can be tuned by varying the orientation of the magnetic vector. Specifically, we find an AHC of approximately 400 S/cm when the magnetization is aligned along the $c$-axis. Overall, our study unveils a rich interplay of pressure and altermagnetism in the model MnTe system, while introducing orthorhombic MnTe as a candidate altermagnet.

\section{Computational Details}

Our density functional theory (DFT) calculations were performed using the Quantum Espresso package~\cite{giannozzi2009quantum}. We used ultrasoft pseudopotentials and the Perdew-Burke-Ernzerhof (PBE) functional within the generalized gradient approximation (GGA)~\cite{perdew1996generalized}. The kinetic energy cut-off for the wavefunction was taken as 60 Ry. To determine the equilibrium lattice parameters of $\alpha$-MnTe and $\gamma$-MnTe variable-cell relaxation calculations were performed with a $k$-grid of $5 \times 5 \times 3$ and $5 \times 8 \times 5$, respectively. For the density of states (DOS) calculations, a denser $15 \times 15 \times 9$ and $15 \times 16 \times 15$ $k$-grids were employed for $\alpha$-MnTe and $\gamma$-MnTe, respectively. The convergence threshold for the self-consistent field was taken as 10$^{-9}$ Ry/atom. To correct the exchange–correlation of Mn $d$-electrons, the DFT+$U$ approximation was applied~\cite{cococcioni2005linear}. Specifically, $U=3$ eV was chosen for Mn $d$ orbitals in the calculations to match previous theoretical and experimental results~\cite{betancourt2023spontaneous,rooj2024hexagonal,szuszkiewicz2006spin}.

Relativistic band dispersion and anomalous Hall conductivity were calculated by incorporating spin-orbit coupling (SOC) within the GGA+$U$ scheme. We utilized maximally localized Wannier functions obtained from the Wannier90 code~\cite{mostofi2014program} to construct a tight-binding Hamiltonian. Using this, we then calculated the anomalous Hall conductivity employing the WannierBerri code~\cite{tsirkin2021high}. \textcolor{black}{To calculate AHC, we employed a $k$-mesh of 55 $\times$ 55 $\times$ 30 for $\alpha$-MnTe and 30 $\times$ 48 $\times$ 27 for $\gamma$-MnTe.} Magnetic space groups were determined using the amcheck program~\cite{smolyanyuk2024tool}.

\section{Results and Discussion}

\begin{figure}[t]
\centerline{\includegraphics[scale=.325]{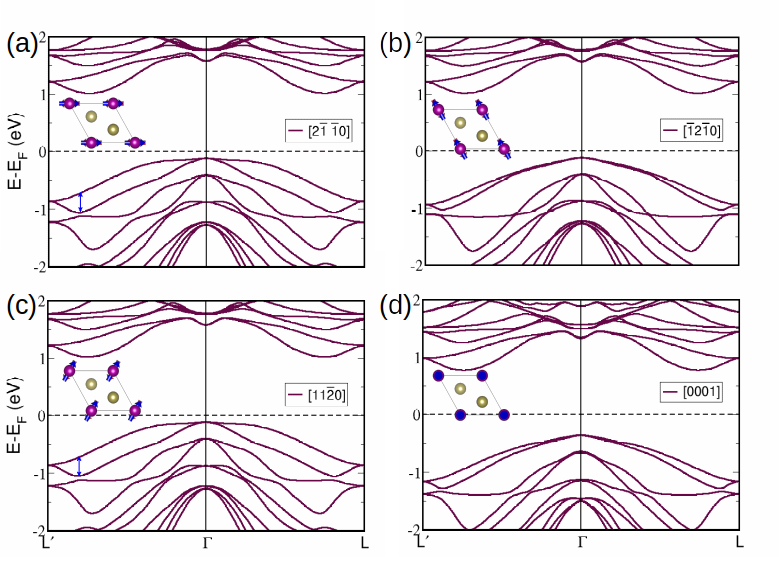}}
\caption{\textbf{Relativistic band dispersion of $\alpha$-MnTe and effect of magnetic moment orientations.} Relativistic band dispersion of $\alpha$-MnTe along the $L'-\Gamma-L$ path when the magnetic moments are aligned along [2$\overline{1}$$\overline{1}$0], [$\overline{1}$2$\overline{1}$0], [11$\overline{2}$0], and [0001], respectively, are shown in (a)-(d), respectively. Corresponding magnetic configurations in the $ab$ plane are
illustrated in the insets. The relativistic band splitting for the bands at valence band edge exceeds 0.30 eV when the magnetization is along [2$\overline{1}$$\overline{1}$0] and [11$\overline{2}$0], which are marked in blue arrows.}
\label{fig_Hex_SOC}
\end{figure}

\subsection{Crystal structure, symmetries and electronic structure}

$\alpha$-MnTe crystallizes in the NiAs-type structure (space group: $P6_3/mmc$, 194)~\cite{d2005low,krause2013structural,wang2022concurrent}, with alternating Mn and Te planes. These planes display a simple hexagonal Mn lattice and a hexagonal close-packed Te lattice with AB stacking, as shown in Fig.~\ref{fig_MnTe_hex_plots} (a). Mn atoms occupy the $2a$ Wyckoff position, which has a site symmetry of $\overline{3}m$, and Te atoms are at the $2c$ Wyckoff position, with a site symmetry $\overline{6}m2$. The configuration with antiparallel spin orientations is energetically favorable compared to the state with parallel spin alignment, with an energy difference of 0.55 eV/f.u.  The geometrically optimized lattice parameters for $\alpha$-MnTe are $a=b=4.21$ \AA{} and $c=6.70$ \AA{}, which closely match experimental measurements~\cite{d2005low}. In this structure, Mn atoms are ferromagnetically ordered within planes that are antiferromagnetically stacked along the $c$-direction. The magnetic moment per Mn atom is calculated to be 4.48 $\mu_B$, which is slightly below the range obtained from neutron diffraction experiments (4.7-5 $\mu_B$)~\cite{kriegner2017magnetic}. The net magnetization vanishes, as each unit cell contains two Mn atoms with identical magnetic moments but opposite orientations.

We begin by examining the non-relativistic electronic band structure, which reveals spin-splitting along the high-symmetry $L'-\Gamma-L$ path [Fig.~\ref{fig_MnTe_hex_plots}(b)]. In contrast, no spin-splitting is observed along the conventional high symmetry path $L-A-\Gamma-M-K-H-A$~\cite{betancourt2023spontaneous,osumi2024observation,rooj2024hexagonal}. Unlike in ferromagnetic materials, the spin-splitting along $L'-\Gamma-L$ demonstrates momentum dependence, with opposite sign spin-splittings for opposite momenta. The projected DOS reveals that the valence band states are predominantly composed of Te 5$p$ orbitals, while the conduction band is primarily made up of Mn 3$d$ orbitals [Fig.~\ref{fig_MnTe_hex_plots}(c)]. Additionally, the total DOS for spin-up and spin-down states are equal, providing further confirmation of zero magnetization.

The origin of non-relativistic spin-splitting in the $\alpha$-MnTe structure is attributed to the breaking of time-reversal symmetry~\cite{betancourt2023spontaneous}. While Mn atoms with opposing alignments are connected by translation or inversion, the opposite-spin sublattices are not linked by crystallographic translation or inversion transformations. The placement of Te atoms, forming octahedra around the Mn atoms at non-centrosymmetric positions, disrupts the time-reversal symmetry within the crystal lattice. As a result, the two opposite-spin sublattices are linked by a non-symmorphic six-fold screw axis rotation ($\Tilde{C_{6}}$) and a $C_2$ rotation in the spin space as shown in Fig.~\ref{fig_MnTe_hex_plots}(a)~\cite{betancourt2023spontaneous}. Visualizations of the magnetization density, as shown in Fig.~\ref{fig_MnTe_hex_plots}(e) and constant energy surfaces at $E=E_F-1$ eV, in Fig.~\ref{fig_MnTe_hex_plots}(f), distinctly depict the connection between opposite-spin sublattices via a rotation.

The altermagnetic phase is described by a nontrivial spin Laue group~\cite{vsmejkal2022beyond,vsmejkal2022emerging}

\begin{equation}
     \textbf{R$_s$} =[\mathcal{E}||\textbf{H}]+[C_2||\textbf{G-H}],   
\end{equation}

where ${\mathcal{E}}$ denotes the spin-space identity transformation, and \textbf{H} represents subgroup containing half of the real-space transformations of the nonmagnetic crystallographic group \textbf{G}~\cite{vsmejkal2022beyond,vsmejkal2022emerging}. Here the operation on the left of double vertical bar acts solely on the spin space and the one on the right of the double vertical bar acts solely on the real space. For $\alpha$-MnTe, which belongs to the crystallographic space group $P6_3/mmc $, the crystallographic point group \textbf{G} is $6/mmm$ and the sublattice point group ${H}$ is $\overline{3}m$. $\alpha$-MnTe follows the spin symmetry,

\begin{equation}
     [\mathcal{E}||{\overline{3}m}]+[C_2||G\backslash mmm|t'].
\end{equation}
    
Here $\overline{3}m$ and G$\backslash$$mmm$ include the symmetry operations \{$ \mathcal{E}, P, C^{\pm}_{3z}, C_{2d}, C_{2x}, C_{2y}, M_d, M_x, M_y $\} and \{$C_{2z}, {\pm}C^{\pm}_{6z}, C_{2d_\perp}, C_{2d_{12}}, C_{2d_{21}}, M_z, M_{d_\perp}, M_{d_{12}}, M_{d_{21}}$\}, respectively. The $\hat{d}$, $\hat{d}_\perp$, $\hat{d_{12}}$, and $\hat{d_{21}}$ axes are along $\hat{x}+\hat{y}$, $\hat{x}-\hat{y}$, $2\hat{x}+\hat{y}$, and $\hat{x}+2\hat{y}$, respectively. Here $\mathcal{E}$ is the identity, $P$ is the inversion, $C_{ni}$ is a $n$-fold rotation around the axis $\hat{i}$, and $M_i$ represents a mirror operation with respect to the plane perpendicular to $\hat{i}$. The half translation along the $z$ axis is represented by $t'$.

We next explore the influence of SOC in $\alpha$-MnTe and the interplay of the band structure and the direction of magnetic moments. Fig.~\ref{fig_Hex_SOC}(a)-(d) illustrate the relativistic band dispersion when the magnetic moments are aligned along [2$\overline{1}$$\overline{1}$0], [$\overline{1}$2$\overline{1}$0], [11$\overline{2}$0], and [0001], respectively. The orientation of the magnetic moment profoundly influences the band splitting and dispersion. For instance, when the magnetization vector is oriented along the [2$\overline{1}$$\overline{1}$0] and [11$\overline{2}$0] directions, the valence band edge at $\Gamma$ exhibits a nearly flat profile, in comparison to the other configurations. Additionally, the degree of band splitting also varies with the magnetization orientation. When the magnetization is along [2$\overline{1}$$\overline{1}$0] and [11$\overline{2}$0], the splitting of the bands near the valence band edge exceeds 0.30 eV, whereas for  [$\overline{1}$2$\overline{1}$0] direction, the band splitting is less than half of the former cases. Our results suggest a direct way to control the spin splitting by means of the magnetic moment orientations.

\begin{centering}
\begin{table*}
\caption{\textbf {Symmetries and non-zero AHC components of $\alpha$-MnTe structures for various magnetization directions.} The magnetic space group, symmetry operations, and non-zero components of AHC present in the $\alpha$-MnTe for different magnetic orientation directions. Translation vectors, $t_1$, $t_2$, and $t_3$ are ($0,0,\frac{1}{2}$), ($\frac{1}{2}, \frac{1}{2}, 0$) and ($\frac{1}{2}, \frac{1}{2}, \frac{1}{2}$), respectively. The $\hat{d}$, $\hat{d}_\perp$, $\hat{d_{12}}$, and $\hat{d_{21}}$ axes are along $\hat{x}$+$\hat{y}$, $\hat{x}$-$\hat{y}$, $2\hat{x}$+$\hat{y}$, and $\hat{x}$+$2\hat{y}$, respectively. }
\label{Table_sym_hex}

\begin{tabular}{c c c c}
\hline
\hline
Magnetization direction & Magnetic space group & Symmetries & Non-zero component of AHC\\
\hline
\hline
\\
\multirow{2}*{along $a$} & \multirow{2}*{$Cmcm (\#63.457) $ } & $\mathcal{E}, P,$  & \multirow{2}*{None}  \\
& & $C_{2x}, \{C_{2y}|t_1\}, \{C_{2z}|t_1\}, $ \\&&$\{C_{2x}|t_2\}, \{C_{2y}|t_3\}, \{C_{2z}|t_3\} $,\\
&&$M_x, \{M_y|t_1\}, \{M_z|t_1\}$ \\
&& $\{M_z|t_2\}, \{M_y|t_3\}, \{M_z|t_3\}$  \\

\multirow{2}*{along $c$} & \multirow{2}*{$P6_3'/m'm'c(\#194.268)  $ } & $\mathcal{E}, P,$  & \multirow{2}*{None}  \\
& & $  C^{\pm}_{3z},  -C^{\pm}_{3z}, $ 
\\&& $  \{C_{2d_\perp}|t_1\}, C_{2d_{12}}|t_1, \{C_{2d_{21}}|t_1\} $ \\ && $ \{M_{d_\perp}|t_1\}, \{M_{d_{12}}|t_1\}, \{M_{d_{21}}|t_1\},$ \\ &&
$  \{{\pm}\tau C^{\pm}_{6z}|t_1\}$, \\ &&
$\{\tau C_{2z}|t_1\}, \tau C_{2x}, \tau C_{2d}, \tau C_{2y}, $ \\ &&
$\{\tau M_{z}|t_1\}, \tau M_{x}, \tau M_{d}, \tau M_{y}, $ 
\\

\multirow{2}*{30\degree ~from $a$} & \multirow{2}*{ $ Cm'c'm(\#63.462) $ } & $\mathcal{E}, P,$  & \multirow{2}*{$\sigma_{xy}$}  \\
& & $ \{C_{2z}|t_1\}, \tau C_{2x}, \{\tau C_{2y}|t_1\} $ \\&& $ \{M_z|t_1\}, \tau M_x, \{\tau M_y|t_1\}, $ &
\\

\hline
\hline
\end{tabular}
 \end{table*}
 \end{centering}
\subsection{Anomalous Hall conductivity and symmetries}

As we noted, AHE is a phenomenon that arises due to broken time-reversal symmetry and strong spin-orbit coupling~\cite{nagaosa2010anomalous}. Additionally, its manifestation is strongly dependent on the relativistic magnetic symmetries which depends on the orientation of the magnetic-order vector with respect to the crystal axes~\cite{betancourt2023spontaneous}. There is a violation of time-reversal symmetry in $\alpha$-MnTe, and we also found that its electronic structure is significantly influenced by the direction of magnetization. Motivated by these factors, we next initiate an investigation into the AHE in $\alpha$-MnTe under different magnetization conditions. \textcolor{black}{Our calculations indicate that the easy axis of magnetization for $\alpha$-MnTe is in-plane at lower pressure values. As pressure increases, however, the easy axis shifts to the out-of-plane direction.}

\begin{figure}[t]
\centerline{\includegraphics[scale=.35]{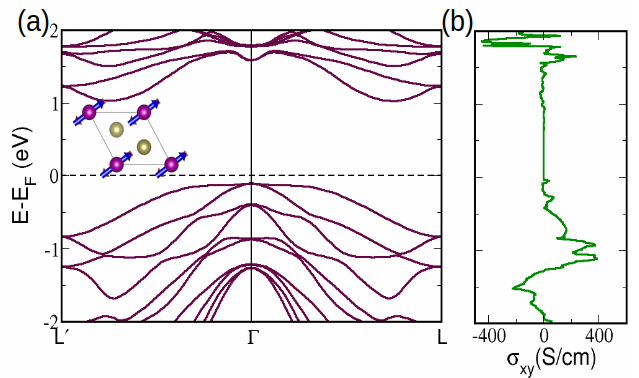}}
\caption{\textbf{Relativistic band structure and anomalous Hall conductivity in $\alpha$-MnTe.} The electronic band structure of $\alpha$-MnTe when the magnetization vector aligns 30$\degree$ from $a$-axis is shown in (a). Here, zero energy corresponds to the Fermi energy. Corresponding magnetic configuration in the $ab$ plane is illustrated in the inset. The anomalous Hall conductivity as a function of $E-E_F$ is shown in (b). The anomalous Hall conductivity is remarkably high, reaching more than 370 S/cm for states at $E=E_F-0.9$ eV.}
\label{fig_Hex_AHC}
\end{figure}

First, we investigate the AHE when the magnetization vector is aligned along the $a$ ([2$\overline{1}$$\overline{1}$0]), $b$ ([$\overline{1}$2$\overline{1}$0]), and $c$ ([0001]) axes. AHE vanishes when the magnetization vector aligns with these crystal axes. Next, we consider magnetization passing through the center of the basal plane, i.e., along [11$\overline{2}$0]. In this case as well, no AHE is observed. Subsequently, we orient the magnetization vector at a 30$\degree$ angle from the $a$-axis in the $ab$ plane. Here, we found the $xy$ component of the Hall conductivity ($\sigma_{xy}$) to be finite. This indicates that the Hall current is in the $x$ direction, with the applied electric field along the $y$ direction, and the Hall vector $h$ parallel to the $c$-axis. The $x$ and $z$ directions are parallel to the $a$-axis and $c$-axis of the crystal, respectively whereas $y$ direction is at 90$\degree$ from $a$-axis. The $z$-direction is parallel to $c$-axis of the crystal. The Hall vector $h$ remains perpendicular to the in-plane magnetization vector. Figure~\ref{fig_Hex_AHC} illustrates the AHC component, $\sigma_{xy}$, as a function of $E-E_F$. We find that the AHC exceeds 370 S/cm for states in the valence band approximately 0.9 eV below the Fermi level. However, it decreases near the Fermi level to approximately 35 S/cm below 0.05 eV from the Fermi level and becomes zero at the Fermi level. These observations closely align with a previous study on the same system~\cite{betancourt2023spontaneous}. 
\textcolor{black}{The AHC values that we predict here are large in comparison to the AHC observed in other materials having compensated magnetic structure~\cite{vsmejkal2022anomalous,feng2020observation, kiyohara2016giant, tenasini2020giant, suzuki2016large}.} 

Let us understand our observations in terms of symmetries of the underlying magnetic structures as the magnetization vectors change direction. When the magnetization aligns along the $a$-axis, the magnetic space group symmetry of $\alpha$-MnTe is $cmcm (\#63.457)$, exhibiting rotation symmetries $C_{2x}$, \{$C_{2y}|t$\}, \{$C_{2z}|t$\}, \{$C_{2x}|t'$\}, \{$C_{2y}|t''$\}, and \{$C_{2z}|t''$\}, where $t$, $t'$, and $t''$ are lattice translation vectors along (0,~0,~$\frac{1}{2}$), (0,~$\frac{1}{2}$,~$\frac{1}{2}$), and ($\frac{1}{2}$,~$\frac{1}{2}$,~$\frac{1}{2}$) directions, respectively. It also possesses a mirror plane $M$ perpendicular to its magnetization direction. These symmetries prevent the existence of AHE and as a results we find that all components of AHC vanish. When the magnetization vector aligns at 30$\degree$ from the $a$-axis, the magnetic space group symmetry changes to $Cm'c'm (\#63.462)$, which follows different symmetry operations detailed in Table~\ref{Table_sym_hex}. \textcolor{black} {In this case, the system hosts two-fold screw rotation along the $c$-axis ($C_{2z}$), along with two other in-plane two-fold rotations combined with time-reversal operation ($\tau$) perpendicular to $c$. A Hall pseudovector can exist along the $c$ axis because the orthogonal rotation axes, combined with the time reversal operation ($\tau C_{2x}$, $\tau C_{2y}$), leave it invariant. Other components of the pseudovector must vanish due to the presence of the two-fold screw axis along the $c$ direction~\cite{betancourt2023spontaneous}.} By performing further rotations of the magnetization vector in the $ab$ plane, we discover that high values of AHC are obtained when the magnetization vector aligns at angles of 90$\degree$, 150$\degree$, 210$\degree$, 270$\degree$ and 330$\degree$. Among these, the maximum AHC components at 30$\degree$, 150$\degree$, and 270$\degree$ exhibit a positive value of $\sigma_{xy}$, whereas at 90$\degree$, 210$\degree$, and 330$\degree$, they take negative $\sigma_{xy}$ values. The magnetic space group symmetry in these cases remains the same as when the magnetization vector is at a 30$\degree$ angle from the $a$-axis, i.e., $Cm'c'm (\#63.462)$. The AHC exhibits a periodic nature and vanishes at every 60$\degree$ angle from the $a$-axis, corresponding to the symmetry of the structure. The magnetic space group of the structure remains unchanged when the magnetization vector changes by an angle of 60$\degree$ in the $ab$-plane. Thus, our calculations show that the magnetization vector orientation is a facile tool to tune the AHC in this class of altermagnets.

\subsection{Effects of pressure}

Experimental studies on MnTe reveal that its hexagonal phase is favorable up to a pressure value of $\approx$20 GPa, beyond which a transition to a different orthorhombic structure occurs~\cite{mimasaka1987pressure,wang2022concurrent}. Therefore, we next conduct a systematic study from 0 GPa to 30 GPa, to comprehensively understand the effects of pressure on the system. We will analyze the effects of pressure on both the hexagonal as well as the orthorhombic phases by means of detailed density functional calculations.

\begin{figure}[t]
\centerline{\includegraphics[scale=.35]{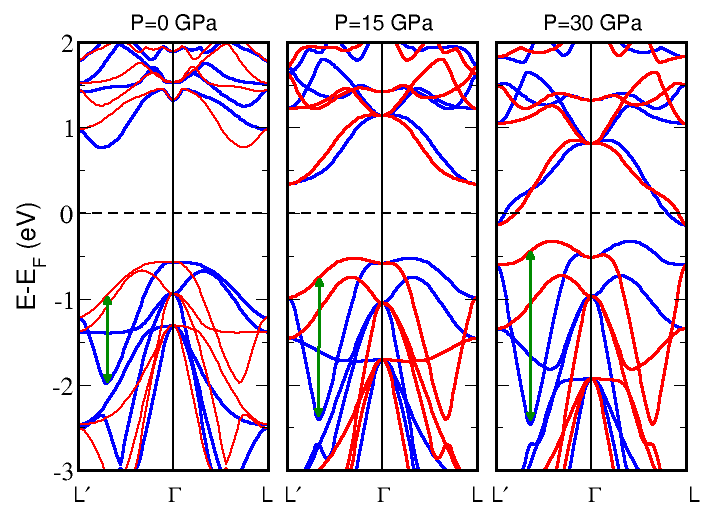}}
\caption{\textbf{The spin-split band structure of $\alpha$-MnTe with pressure.} The spin-split band structure of $\alpha$-MnTe along the $L'-\Gamma-L$ path at various pressures. Spin-up and spin-down bands are represented by blue and red solid lines, respectively. Pressure significantly changes the maximum spin-splitting which is indicated by a green arrow. At low pressures, spin-splitting is nearly 1 eV. As the pressure increases, spin-splitting also increases, nearly doubling at 30 GPa.}
\label{fig_Hex_pressure_band}
\end{figure}

Let us begin with the effects of pressure on $\alpha$-MnTe. We find that applied pressure reduces the lattice parameters and bond length significantly. The magnetic moment for Mn atoms decreases with pressure, which is consistent with experimental observations~\cite{wang2022concurrent} (see Appendix~\ref{SI-press-hex_lattice}). We discover that these changes in $\alpha$-MnTe under pressure result in significant
alterations in the electronic structure as well (Fig.~\ref{fig_Hex_pressure_band}). One of the most intriguing findings from our band structure analysis under varying pressure conditions is the persistence of spin-split bands even at very high pressures, indicating the robustness of altermagnetism under extreme pressure conditions. Furthermore, significant changes occur in the band structure, including spin-splitting, band curvature, and band crossing. The maximum spin-splitting at the valence band side nearly doubles to 2 eV from the value of 1.05 eV as pressure varies increases from 0 to 30 GPa. This enhancement arises as each band within the pair of altermagnetic spin-split bands (indicated by a green arrow in Fig.~\ref{fig_Hex_pressure_band}) shifts their energy level in the opposite manner relative to the Fermi level -- one band moves closer to the Fermi level while the other moves away as pressure increases. Conversely, a completely different scenario is observed in the conduction band, where spin-splitting decreases with increasing pressure. This relatively opposite phenomenon in valence and conduction bands can be understood in terms of orbital hybridization. From an analysis of the PDOS, we observe that the valence bands are predominantly contributed by Te $p$ orbitals. However, as pressure increases, the contribution from Mn $d$ orbitals become comparable to that from Te $p$ orbitals (see Appendix~\ref{SI-press-dos-band}). This increase in the hybridization can be attributed to reduction in bond length between Mn and Te atoms under pressure. On the other hand, the conduction bands are mainly contributed by Mn $d$ orbitals -- Mn2 and Mn1 contributing to the spin up bands and down bands, respectively. Increasing pressure brings Mn atoms closer, narrowing the energy difference between Mn1 and Mn2 orbitals and, thus, reducing the spin-splitting.

Other notable changes that occur in the electronic structure due to pressure are the changes in curvature of bands and position of band edges. At low pressures, the valence band maximum occurs at the $\Gamma$ point, exhibiting a nearly flat dispersion. However, as pressure increases, the valence band maximum shifts away from $\Gamma$, and it becomes more dispersive. Conduction band edges located between $\Gamma$-L are shifted to the high symmetric point $L$ at high pressures approaching 30 GPa.

The substantial spin-splitting observed in $\alpha$-MnTe has already positioned it as a prototypical altermagnet as well as a promising material for spintronic applications. Our study highlights pressure as a significant tuning parameter for enhancing the spin-splitting and altermagnetic properties of $\alpha$-MnTe.

\begin{figure*}[t]
\centerline{\includegraphics[scale=.425]{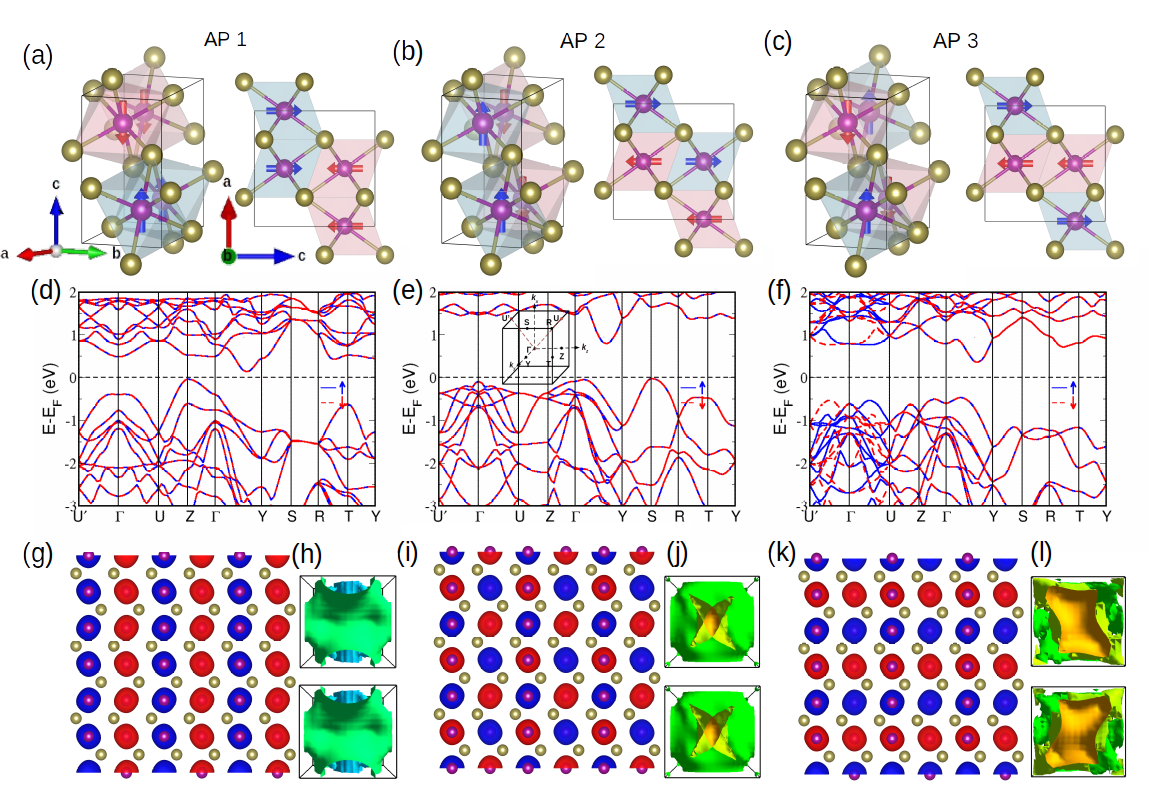}}
\caption{\textbf{Crystal and electronic structure of orthorhombic $\gamma$-MnTe.} Three possible antiparallel (AP) spin arrangements of $\gamma$-MnTe are illustrated in (a)-(c). Corresponding spin-polarized band structures for AP1, AP2 and AP3 are depicted in panels (d), (e), and (f) respectively. The blue solid lines represent spin-up bands, while red dashed lines represent spin-down bands. Notably, for AP1 and AP2, spin-up and spin-down bands are degenerate throughout the entire path, whereas spin-split bands are evident for AP3 along the $k$-path $U'-\Gamma-U$.
The magnetization density distributions for AP1, AP2, and AP3 are visualized in panels (g), (i), and (k) respectively. The constant energy surfaces at $E=E_F-1$ eV are shown in panels (h), (j), and (l) for AP1, AP2, and AP3, respectively. Top panels are for spin-up and bottom panels are for spin-down. The spin-split band structure, magnetization density, and constant energy surfaces of the opposite-spin sublattices, which are connected by rotation indicate that AP3 configuration of $\gamma$-MnTe is an altermagnet, whereas AP1 and AP2 are antiferromagnets.}
\label{fig_Ortho_AFM_structures}
\end{figure*}

\begin{centering}
\begin{table*}[t]
\caption{\textbf{Comparison of lattice parameters and energetics of different magnetic phases of $\gamma$-MnTe.} The lattice constants $a$, $b$, and $c$, volume of the unit cell, magnetic moment of Mn, energy per MnTe, and magnetic space group symmetry for various magnetic structures of $\gamma$-MnTe, when the magnetic moment orientation is along $z$-axis. AP1, AP2 and AP3 represent the $\gamma$-MnTe structures with antiparallel spin alignment, as depicted in Fig.~\ref{fig_Ortho_AFM_structures} (a-c), respectively. Parallel denotes the $\gamma$-MnTe structure with parallel spin alignment.}
\centering
\begin{tabular}{c c c c c c c c}
\hline
\hline
Spin arrangement & $a$ (\AA) & $b$ (\AA) & $c$ (\AA) & $V$ (\AA$^3$) & $\mu_{Mn}$ ($\mu_B)$ & $E$/f.u. (eV) & Magnetic space group symmetry \\
\hline
\hline
AP1 & 6.7881& 4.2068 & 7.3089 & 208.7209&4.52& -3296.8574 & $Pnm'a$(\#62.444)\\
AP2 & 6.7443 & 4.2021 & 7.3244 & 207.5801& 4.53& -3296.8941 & $Pn'm'a'$(\#62.449)\\
AP3 & 6.7235 & 4.2175 & 7.2736 & 206.2528& 4.49& -3296.9140 & $Pnm'a'$(\#62.447) \\
Parallel & 5.9042 & 4.3529  & 7.5105 & 193.0296 & 4.06 & -3296.4142 & $Pn'm'a$(\#62.446)\\
\hline
\hline
\end{tabular}
\label{Table_ortho_lattice_parameters}
 \end{table*}
\end{centering}

\subsection{Orthorhombic MnTe}

After analyzing the hexagonal phase of MnTe, we further extend our investigation to include the orthorhombic phase of MnTe, which is also known as $\gamma$-MnTe. As we noted, the phase transition of MnTe from hexagonal phase to the orthorhombic phase under pressure has been demonstrated~\cite{mimasaka1987pressure,wang2022concurrent}. Given the lack of detailed studies on $\gamma$-MnTe, we next conduct a comprehensive analysis of this MnTe structure.

$\gamma$-MnTe crystallizes in the MnP-type structure (space group: Pnma, 62)~\cite{wang2022concurrent}. In this structure, the Mn atom occupies the Wyckoff position 4c with site symmetry group $.m.$~\cite{aroyo2011crystallography}. The magnetic arrangement of Mn atoms over the possible Wyckoff sites can lead to changes in the magnetic space group of the structure, which in turn may result in changes of the material properties. Therefore, we analyze various magnetic configurations of $\gamma$-MnTe, including both parallel and antiparallel (AP) spin arrangements. The collinear AP spin arrangement manifests in three distinct configurations, depicted in Fig.~\ref{fig_Ortho_AFM_structures} (a)-(c), we name them as AP1, AP2, and AP3 phases. In AP1, the magnetic moments of Mn ions align parallel to each other within the hexagonal basal plane, while they align antiparallel along the $c$-direction. In AP2, Mn ions are aligned antiparallel to each other in the basal plane and along the $c$-direction. In AP3, Mn ions are aligned antiparallel in the basal plane but parallel along the $c$-direction. We conducted geometric optimizations for all these structures, and various parameters were analyzed. The resulting lattice parameters, magnetic properties and energetics are summarized in Table~\ref{Table_ortho_lattice_parameters}. Minor variations were observed in the lattice parameters and magnetic moments of Mn across different configurations. Notably, the AP configurations exhibit significantly lower energies than the structure with parallel spin arrangement. In comparison with the parallel spin configuration structure, AP1, AP2, and AP3 spin configurations exhibit lower energy per formula unit (f.u.) by 0.4432 eV, 0.4799 eV, and 0.4998 eV, respectively (see Table~\ref{Table_ortho_lattice_parameters}), with AP3 being the most stable.

Upon a careful analysis of of the band structures of all the $\gamma$-MnTe configurations with AP spin arrangements [Fig.~\ref{fig_Ortho_AFM_structures}(d)-(f)], we found significant distinctions in their properties, despite all configurations exhibiting a net magnetization of zero. The net magnetization becomes zero when the opposite-spin sublattices are connected by either a lattice translation, spatial inversion, or rotation~\cite{smolyanyuk2024tool}. Let us recall that if the connection occurs through a lattice translation or spatial inversion, the material is classified as antiferromagnetic; otherwise, it is categorized as an altermagnet~\cite{vsmejkal2022beyond,vsmejkal2022emerging,smolyanyuk2024tool}. Therefore, we placed particular emphasis on scrutinizing the magnetic symmetries and specifically the connections between opposite-spin sublattices in all these configurations. Magnetic configurations of AP1 and AP2 show that, opposite-spin sublattices in these are connected by a spatial inversion. So, these materials come under the class of antiferromagnets. In contrast, the opposite-spin sublattices of AP3 are not connected by spatial inversion or lattice translation. This deviation is attributed to the specific arrangement of Mn atoms and the presence of Te atoms in noncentrosymmetric positions, which prevents the opposite-spin sublattices from connecting via inversion or translation. Instead, they are linked by a two-fold screw rotation, thus falling into the magnetic class of altermagnets. We further note that AP3 is found to be energetically the most stable one among these AP phases from our DFT calculation, as summarized in Table~\ref{Table_ortho_lattice_parameters}.

\begin{figure}[t]
\centerline{\includegraphics[scale=.3]{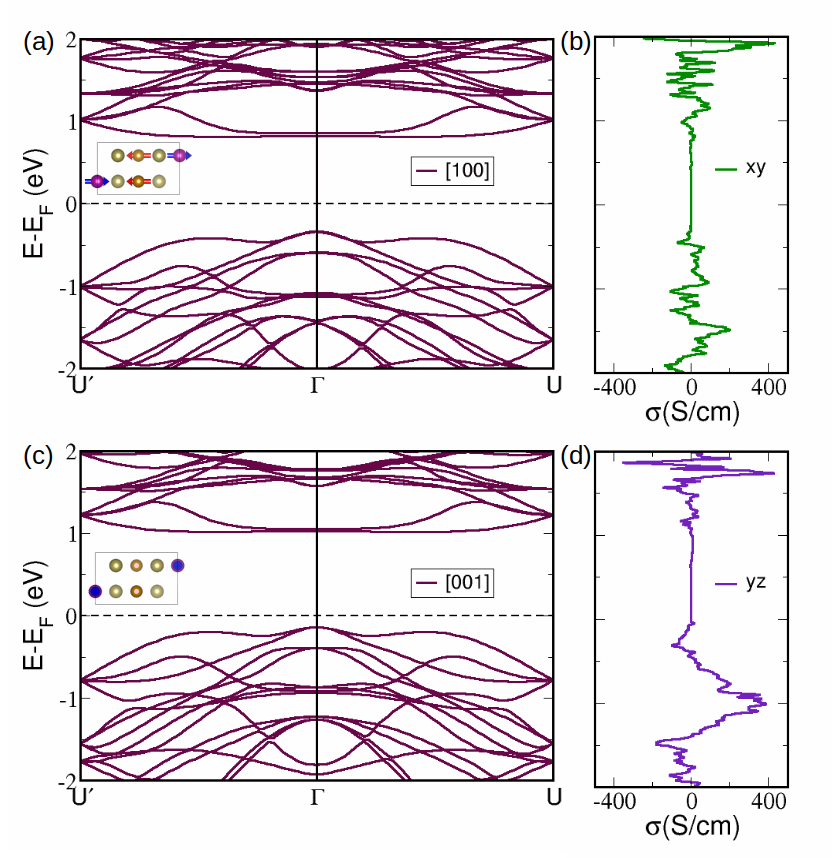}}
\caption{\textbf{Relativistic band structure and anomalous Hall conductivity of $\gamma$-MnTe.} The relativistic band structures of $\gamma$-MnTe when the magnetization is along [100] and [001] are shown in (a) and (c), respectively. The corresponding AHC components calculated for these magnetizations are shown in (b) and (d). AHC components depend sensitively on the magnetization direction. A high value of AHC, nearly 400 S/cm, occurs  when the magnetization is along [001].}
\label{fig_Ortho_SOC_AHC}
\end{figure}

We present the spin-polarized band structures along the $k$-path $U'-\Gamma-U-Z-\Gamma-Y-S-R-T-Y$ for all the AP phases in Fig.~\ref{fig_Ortho_AFM_structures}(d)-(f). These band structures exhibit significant differences, including the variation in band gap. AP1 phase demonstrates a lower band gap compared to the other phases, with a value of 0.18 eV. AP2 and AP3 phases, on the other hand, have comparable band gaps of 0.81 eV and 0.83 eV, respectively. An intriguing aspect revealed by the band structures is their indication of the corresponding magnetic classes. In AP1 and AP2 phases of $\gamma$-MnTe, both spin-up and spin-down bands remain degenerate throughout the $k$-path, suggesting their antiferromagnetic nature. Conversely, in AP3, degeneracy occurs solely along the $U-Z-\Gamma-Y-S-R-T-Y$ path. Remarkably, we find a momentum-dependent spin-splitting along the $U'-\Gamma-U$, with the direction of spin-splitting reversing upon reversing the momentum path. These characteristics are indicative of altermagnetic materials~\cite{vsmejkal2022emerging,vsmejkal2022beyond}. From our calculated magnetization density plots shown in Fig.~\ref{fig_Ortho_AFM_structures} (g) and (i), it is clear that in the phases AP1 and AP2 of $\gamma$-MnTe, the opposite magnetization densities can be connected by a translation or inversion, which is a typical feature of antiferromagnets. However, in AP3, they are related only through a rotation [Fig.~\ref{fig_Ortho_AFM_structures} (k)], revealing its altermagnetic nature. As expected from the magnetization density plots, the constant energy surfaces are identical for spin-up and spin-down states for AP1 and AP2 phases of $\gamma$-MnTe [Fig.~\ref{fig_Ortho_AFM_structures} (h) and (j)]. However, in altermagnetic AP3 phase, they are connected by a rotation [Fig.~\ref{fig_Ortho_AFM_structures} (l)]. This altermagnetic $\gamma$-MnTe follows a bulk spin-momentum locking and has a $d$-wave symmetry~\cite{vsmejkal2022beyond}. As such, our study presents orthorhombic $\gamma$-MnTe as an unexplored altermagnet candidate.
 
Next, we investigate the relativistic band structures as well as AHE for this $\gamma$ phase. As we have seen in $\alpha$-MnTe earlier, in the present case also the magnetic point group symmetry changes with the magnetization direction (Table~\ref{Table_sym_ortho}). However, unlike $\alpha$-MnTe, we find a variation of the non-zero AHC components with the change in magnetization direction. Our findings are summarized in Table~\ref{Table_sym_ortho}. When the magnetization is along the $a$-axis ([100]), the presence of $C_{2z}$ allows only $\sigma_{xy}$, restricting the $xz$ and $yz$ components to vanish. In a similar way, $C_{2x}$ only allows $\sigma_{yz}$ to be non-zero, when magnetization is along the $c$-axis ([001]). On the contrary, the simultaneous presence of three rotational symmetries along three principle axes prohibits AHE entirely when the magnetization is along the $b$-axis. We also find a change in AHC values from our calculations, when the magnetization direction is varied as can be seen from Fig.~\ref{fig_Ortho_SOC_AHC}. A maximum value of 200 S/cm at $E = E_{F}-1.5$ eV is obtained, when the magnetization is along the $a$ axis, which enhances to 400 S/cm at $E = E_{F}-1$ eV with a change in magnetization direction towards the $c$-axis.

As we discussed, previous experimental studies have shown that the $\alpha$-MnTe phase undergoes a phase transition to the $\gamma$-MnTe phase at higher pressures~\cite{mimasaka1987pressure,wang2022concurrent}. Therefore, it is important to investigate the existence of altermagnetic properties in the $\gamma$-MnTe phase under higher pressures. To this end, we use the energy obtained from DFT calculations for $\alpha$-MnTe and $\gamma$-MnTe to fit the Murnaghan equation of state~\cite{murnaghan1944compressibility,birch1947finite}. The resulting volume-energy curve shows that at lower pressures, both the $\alpha$-MnTe and $\gamma$-MnTe phases have comparable energy within the density functional calculations. However, at higher pressures above 20 GPa, the $\gamma$-MnTe phase becomes more stable, as observed in the previous experiments. Therefore, we next examine the altermagnetic phase of $\gamma$-MnTe up to 30 GPa. \textcolor{black}{We analyzed the energy of three AP magnetic orientations of $\gamma$-MnTe within this pressure range, and the results indicate that the altermagnetic phase of $\gamma$-MnTe remains the lowest energy state among these three AP phases, across the pressure values we considered.}

Similar to $\alpha$-MnTe, the bond lengths, magnetic moment, and band gap of $\gamma$-MnTe decrease under pressure (see Appendix ~\ref{SI-press-hex_lattice}). However, due to the differences in crystal structure, the effects of external pressure vary between the two structures. The magnetic moment of Mn in $\gamma$-MnTe is slightly lower than that in $\alpha$-MnTe at various pressures, which follows the trend observed in experiments~\cite{wang2022concurrent}. While the band gaps of $\alpha$-MnTe and $\gamma$-MnTe are comparable at low pressures, external pressure does not reduce the band gap of $\gamma$-MnTe as significantly as it does for $\alpha$-MnTe. At 30 GPa, the band gap of $\gamma$-MnTe is almost double of $\alpha$-MnTe, being 0.19 eV for $\alpha$-MnTe and 0.41 eV for $\gamma$-MnTe.

\begin{table*}
\caption{\textbf {Symmetries and non-zero AHC components of $\gamma$-MnTe structures with various magnetization directions.} The magnetic space group, symmetry operations, and non-zero components of AHC present in $\gamma$-MnTe for different magnetic orientation directions. Translation vectors, $t_1$, $t_2$, and $t_3$ are ($0,\frac{1}{2},0$), ($\frac{1}{2}, 0, \frac{1}{2}$) and ($\frac{1}{2}, \frac{1}{2}, \frac{1}{2}$), respectively. }
\centering
\begin{tabular}{c c c c}
\hline
\hline
Magnetization direction & Magnetic space group & Symmetry & Non-zero component of AHC\\
\hline
\hline
\\
\multirow{4}*{$[100]$} & \multirow{3}*{$Pn'm'a$ $~(\#62.446)$ } &$ \mathcal{E}, P,$ & \multirow{4}*{$\sigma_{xy}$}  \\
& & \{$ C_{2z}|t_2\}, \{\tau C_{2x}|t_3\}, \{\tau C_{2y}|t_1$\}, \\
& & \{$ M_z|t_2\}, \{\tau M_x|t_3\}, \{\tau M_y|t_1$\}  \\\\

\multirow{4}*{$[010]$} & \multirow{3}*{$Pnma$ ~(\#62.441)} & $\mathcal{E}, P,$ & \multirow{4}*{None} \\
& & \{$C_{2x}|t_3\}, \{C_{2y}|t_1\}, \{C_{2z}|t_2\}, $ & \\
& & $ \{M_x|t_3\}, \{M_y|t_1\}, \{M_z|t_2\}$ & \\\\

\multirow{4}*{$[001]$} & \multirow{4}*{$Pnm'a'$ ~(\#62.447)} & $\mathcal{E}, P, $  & \multirow{4}*{$\sigma_{yz}$} \\
& & $ \{\tau C_{2x}|t_3\}, \{ C_{2y}|t_1\}, \{\tau C_{2z}|t_2\}, $  & \\
& & \{$\tau M_x|t_3\}, \{M_y|t_2\}, \{\tau M_z|t_2\} $ & \\\\
\hline
\hline
\end{tabular}

\label{Table_sym_ortho}
 \end{table*}

\begin{figure}[t]
\centerline{\includegraphics[scale=.35]{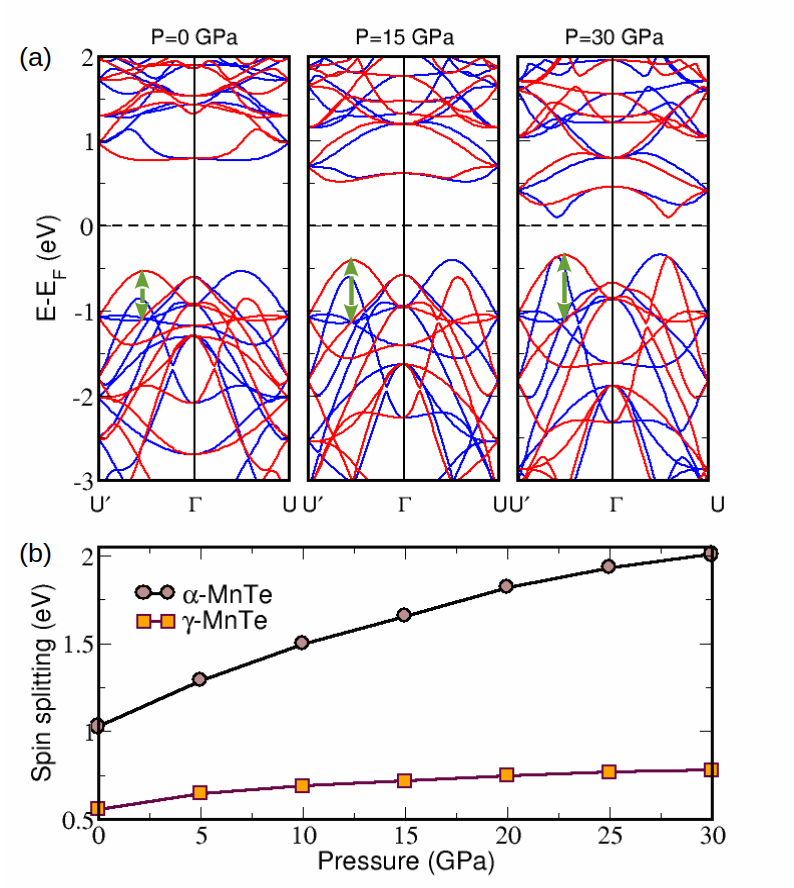}}
\caption{\textbf{Spin-split band structure of $\gamma$-MnTe at different pressure values and spin-splitting as a function of pressure.} (a) The spin-split band structure of $\gamma$-MnTe along the $U'-\Gamma-U$ path is presented at various pressures. Spin-up and spin-down bands are represented by blue and red solid lines, respectively. The maximum spin-splitting in each band structure, indicated by a green arrow, increases with the increase in pressure. The band gap, band dispersion, and the position of conduction band minima are also directly influenced by the applied pressure. (b) The spin-splitting in $\alpha$-MnTe and $\gamma$-MnTe is shown as a function of pressure. The spin-splitting is greater in $\alpha$-MnTe than in $\gamma$-MnTe. In $\alpha$-MnTe, the increase of spin-splitting with pressure is higher compared to that in $\gamma$-MnTe.}
\label{fig_MnTe_ortho_pressure}
\end{figure}

We plot the band structure along the $k$-path $U’-\Gamma-U$, where spin-split bands are obtained at zero pressure, at various pressures as shown in Fig.~\ref{fig_MnTe_ortho_pressure}(a). Most notably, similar to $\alpha$-MnTe, the altermagnetism persists at higher pressure in $\gamma$-MnTe too. Comparing the spin-splitting, a crucial quantitative parameter for altermagnets, between $\gamma$-MnTe and $\alpha$-MnTe is informative, as depicted in Fig.~\ref{fig_MnTe_ortho_pressure}(b) as a function of pressure. We discover that the spin-splitting in $\gamma$-MnTe is nearly half of that in $\alpha$-MnTe under the same pressure conditions. At zero pressure, the spin-splitting in $\gamma$-MnTe is 0.56 eV, while it is 1.03 eV in $\alpha$-MnTe. This distinction in spin-splitting is further underscored when observing the response to pressure for the two structures, as we illustrate in Fig.~\ref{fig_MnTe_ortho_pressure}(b). While both phases exhibit an increase in spin-splitting with pressure, the rate of increase is notably higher in $\alpha$-MnTe than in $\gamma$-MnTe. Specifically, at 30 GPa, the spin-splitting in $\alpha$-MnTe and $\gamma$-MnTe measure 2 eV and 0.78 eV, respectively. This implies that in $\alpha$-MnTe, the spin-splitting nearly doubles when pressure changes from 0 to 30 GPa, whereas in $\gamma$-MnTe, the increase is only approximately 40\%. The projected DOS and band structures of $\gamma$-MnTe (see Appendix~\ref{SI-press-dos-band}) show that, similar to $\alpha$-MnTe, in $\gamma$-MnTe, the valence band is formed by Mn $3d$ and Te $5p$ hybridized orbitals with major contributions from Te $5p$ orbitals. The contribution from Mn $3d$ orbitals increases with pressure. The conduction band is formed from the Mn $3d$ orbitals. These are similar to the case of $\alpha$-MnTe. However, the nature of band structures differs in both cases. Compared to $\alpha$-MnTe, significantly more overlap of bands is present in the valence band of $\gamma$-MnTe and band crossings are also more evident. This may be preventing drastic changes in energy level of bands in the valence band manifold with increasing pressure, resulting in a lower spin-splitting rate. In $\alpha$-MnTe, both bands in the altermagnetic spin-split band pair change their energy levels in opposite directions with respect to the Fermi level at the same rate with pressure ($\approx$0.53 eV). However, in $\gamma$-MnTe, the state at the valence band maximum of the altermagnetic pair changes its energy by approximately 0.2 eV, whereas for the other band in the altermagnetic pair, the change is even less, around 0.03 eV, under the same pressure change. Many other bands present near the lower band of the altermagnetic spin-split band prevents its position from being significantly affected by external pressure. 
As pressure increases, the bands from Mn atoms with opposite magnetization approach each other, leading to a reduction in spin-splitting in the conduction band. It is noteworthy that the sign of spin-splitting in the conduction band is opposite to that at lower pressure. This phenomenon is attributed to the fact that the band forming the conduction band minima at 30 GPa is different from that at lower pressures. Additionally, the momentum difference between the valence band maxima and conduction band minima decreases with pressure.

\section{Summary}

To summarize, in this study, we comprehensively investigated the electronic and magnetic properties of MnTe -- both in the hexagonal ($\alpha$) and orthorhombic ($\gamma$) phases. We demonstrated that $\alpha$-MnTe, a well-known $g$-wave altermagnet, exhibits a remarkably enhanced spin-splitting under external pressure. Specifically, the $\alpha$-MnTe phase exhibits a significant spin-splitting of nearly 1 eV at low pressure, which increases substantially to nearly 2 eV when the pressure is raised to 30 GPa. We also showed that the AHC depends critically on the magnetic moment orientation, suggesting potential tunability. We predicted a high AHC of more than 370 S/cm when the magnetization vector aligns at certain angles in the basal plane and found that the AHC varies periodically with a period of 60$^{\degree}$. In addition to the $\alpha$-MnTe phase, we conducted an in-depth study of $\gamma$-MnTe, which is known to be the more stable phase at higher pressures in experiments. Notably, our results reveal the intriguing discovery of $d$-wave altermagnetism in one of the antiparallel configurations of $\gamma$-MnTe system. Similar to the $\alpha$-MnTe, $\gamma$-MnTe also exhibits larger spin-split bands with pressure, but the degree of the spin-splitting in the bands is almost half that of the former. We showed that $\gamma$-MnTe also exhibits a dependence of AHC on the magnetic moment orientation. However, unlike $\alpha$-MnTe, the direction of the Hall vector also changes with magnetic moment orientation -- giving rise to two components, $\alpha_{xy}$ and $\alpha_{yz}$. Our calculations predict a high AHC of $\approx$ 400 S/cm when the magnetization vector aligns along $c$-axis. 

The findings from our comprehensive study not only open up new research directions focused on the $\gamma$-MnTe phase in the context of altermagnetism, but also suggest the intriguing possibility of tuning band-splitting with pressure and controlling the AHE through the manipulation of magnetization orientation. These insights offer exciting opportunities for the design of devices with customizable electronic and magnetic characteristics.

\textcolor{black}{The findings from our comprehensive study not only open up new research directions focused on the $\gamma$-MnTe phase in the context of altermagnetism, but also demonstrate the stability and robustness of the altermagnetic phase in both the structures of MnTe. This suggests an enhancement the operational range for devices such as magnetic sensors, memory devices, and spin transistors. In addition, our work reveals the intriguing possibility of tuning band-splitting with pressure, paving the way for advanced pressure-sensitive spintronic devices. Furthermore, the ability to control the anomalous Hall effect by manipulating magnetization orientation presents exciting prospects for designing devices with customizable electronic and magnetic properties.}\\

\section*{Acknowledgments}
We thank A. Chakraborty, A. Reja, A. Bandyopadhyay, and N. B. Joseph for the valuable discussions. N.D. would like to acknowledge IoE-IISc Postdoctoral Fellowship. A.B. is supported by Prime Minister's Research Fellowship (PMRF). A. N. acknowledges support from DST CRG grant (CRG/2023/000114).

\bibliography{references}

\appendix

\section{\textcolor{black}{Non-relativistic spin-degenerate nodal surfaces of $\alpha$-MnTe and $\gamma$-MnTe}}
\label{nodalplane_appendix}

\begin{figure}[t]
\centerline{\includegraphics[scale=.4]{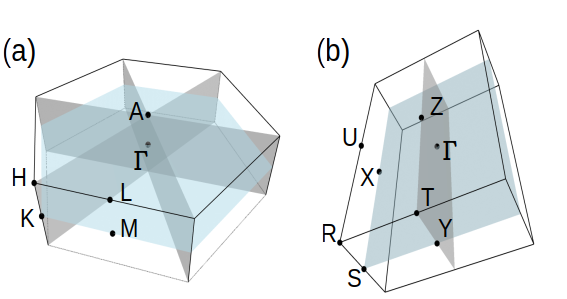}}
\caption{\textbf{\textcolor{black}{Non-relativistic spin-degenerate nodal surfaces of $\alpha$-MnTe and $\gamma$-MnTe.}} \textcolor{black}{The Brillouin zones of $\alpha$-MnTe and $\gamma$-MnTe, with high-symmetry $k$-points marked, are shown in panels (a) and (b). The non-relativistic spin-degenerate nodal surfaces of $\alpha$-MnTe and $\gamma$-MnTe are depicted as blue and gray planes. }}
\label{fig_nodalplane_hex_ortho}
\end{figure}

\textcolor{black}{In Fig.~\ref{fig_nodalplane_hex_ortho} (a) and (b), we show the nodal surfaces within the non-relativistic limit for $\alpha$-MnTe and $\gamma$-MnTe, respectively.
In $\alpha$-MnTe, the spin-degenerate nodal planes are protected by the $[C_2||M_z]$ and $[C_2||C_{6z}]$  symmetries. In $\gamma$-MnTe, the spin-degenerate nodal planes are protected by the $[C_2||M_x]$ and $[C_2||M_z]$ symmetries.}

\section{Details of effects of pressure on structural parameters of $\alpha$-MnTe and $\gamma$-MnTe}
\label{SI-press-hex_lattice}

\begin{figure*}[t]
\centerline{\includegraphics[scale=.32]{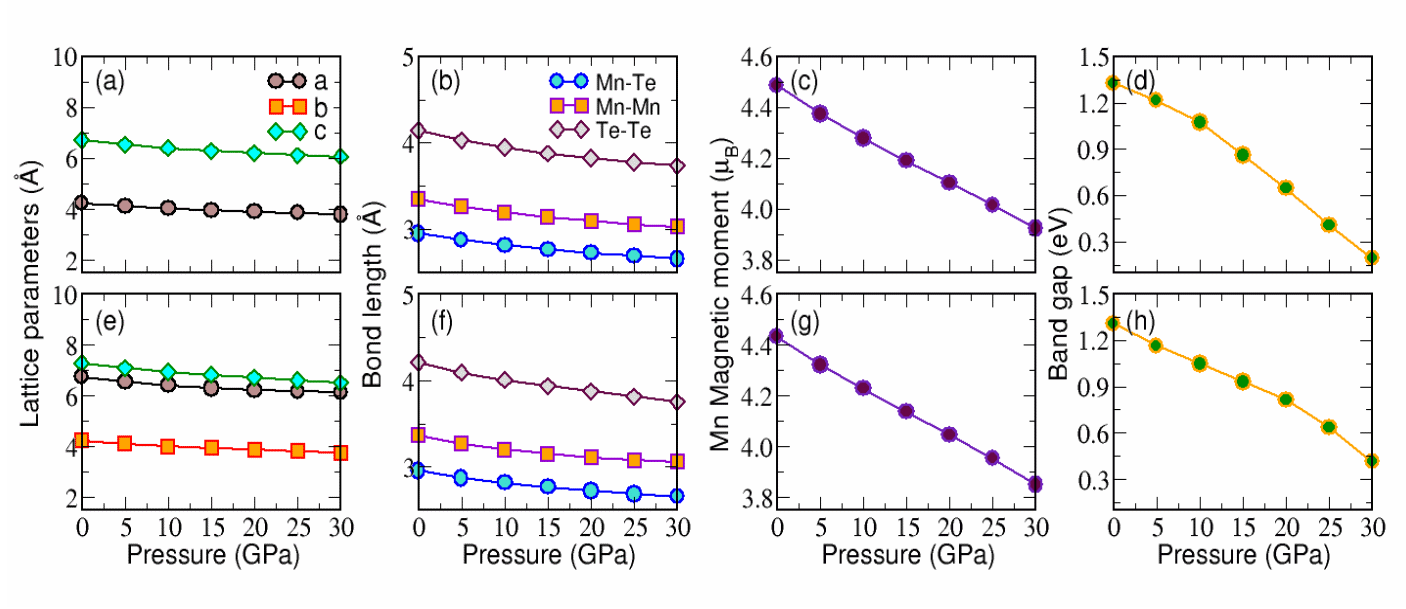}}
\caption{\textbf{The variation in structural parameters of MnTe as a function of pressure.} The lattice parameters, atomic separations, magnetic moment of Mn atoms, and band gap for varying pressure values of $\alpha$-MnTe are shown in the top panels (a)-(d). The bottom panels (e)-(f) depict the variation of the structural parameters same as (a)-(d), for the $\gamma$-MnTe. As a result of applied pressure, a reduction occurs for all these parameters for both phases.}
\label{subfig_press_parameters}
\end{figure*}

The effects of applied pressure on the structural parameters of MnTe are depicted in Fig.~\ref{subfig_press_parameters}, where the top panels (a)-(d) and the bottom panels (e)-(h) show the variation of lattice parameters, bond lengths, Mn magnetic moment, and band gaps for $\alpha$-MnTe and $\gamma$-MnTe, respectively, as a function of applied pressure ranging from 0 GPa to 30 GPa. As a result of applied pressure, lattice parameters and bond lengths decrease for both systems. Under ambient pressure conditions, the lattice constants $a$ and $c$ for the $\alpha$-MnTe system read 4.21 \AA{} and 6.70 \AA{}, respectively, which reduce to 3.78 \AA{} and 6.05 \AA{}, respectively, when the pressure value is tuned to 30 GPa [Fig.~\ref{subfig_press_parameters}(a)]. In a similar way, the lattice parameters, $a$, $b$, and $c$ of $\gamma$-MnTe take up reduced values of 6.12 \AA{}, 3.75 \AA{}, and 6.49 \AA{}, respectively at 30 GPa from 6.72 \AA{}, 4.22 \AA{}, and 7.27 \AA{}, respectively at ambient pressure conditions [Fig.~\ref{subfig_press_parameters}(e)]. A similar trend is followed in bond lengths as well. At 0 GPa, the lengths of Mn-Te, Mn-Mn, and Te-Te bonds in $\alpha$-MnTe are 2.95 \AA{}, 3.35 \AA{}, and 4.14 \AA{}, respectively, which decrease with increasing pressure and reach to 2.66 \AA{}, 3.02 \AA{}, and 3.73 \AA{} at 30 GPa, as illustrated in Fig.~\ref{subfig_press_parameters}(b). The reduction of bond lengths as a function of applied pressure in $\gamma$-MnTe has been depicted in Fig.~\ref{subfig_press_parameters}(f). When pressure changes from 0 GPa to 30 GPa, Mn-Te bond length changes from 2.96 \AA{} to 2.65 \AA{}. Under the same pressure variation, Mn-Mn bond length changes from 3.36 \AA{} to 3.06 \AA{} and Te-Te bond length changes from 4.20 \AA{} to 3.75 \AA{}.

Despite the applied pressure, the net magnetic moment remains zero and the magnetic moment of each Mn decreases with increasing pressure in both the structures, as shown in Fig.~\ref{subfig_press_parameters}(c) and (g). For $\alpha$-MnTe, it changes from 4.48 $\mu_B$ to 3.92 $\mu_B$, under an applied pressure of 30 GPa. Under the same pressure variation, it reduces from 4.43 $\mu_B$ to 3.85 $\mu_B$ for $\gamma$-MnTe. The variation in energy band gap for $\alpha$-MnTe and $\gamma$-MnTe are shown in Fig.~\ref{subfig_press_parameters}(d) and (h). The energy band gap of $\alpha$-MnTe and $\gamma$-MnTe are comparable at 0 GPa, being 1.33 eV and 1.31 eV, respectively. However, band gap reduces more rapidly in $\alpha$-MnTe than $\gamma$-MnTe under pressure. At 30 GPa, it is 0.19 eV for $\alpha$-MnTe, while for $\gamma$-MnTe, it is nearly double at 0.41 eV.

\begin{figure*}
\centerline{\includegraphics[scale=.58]{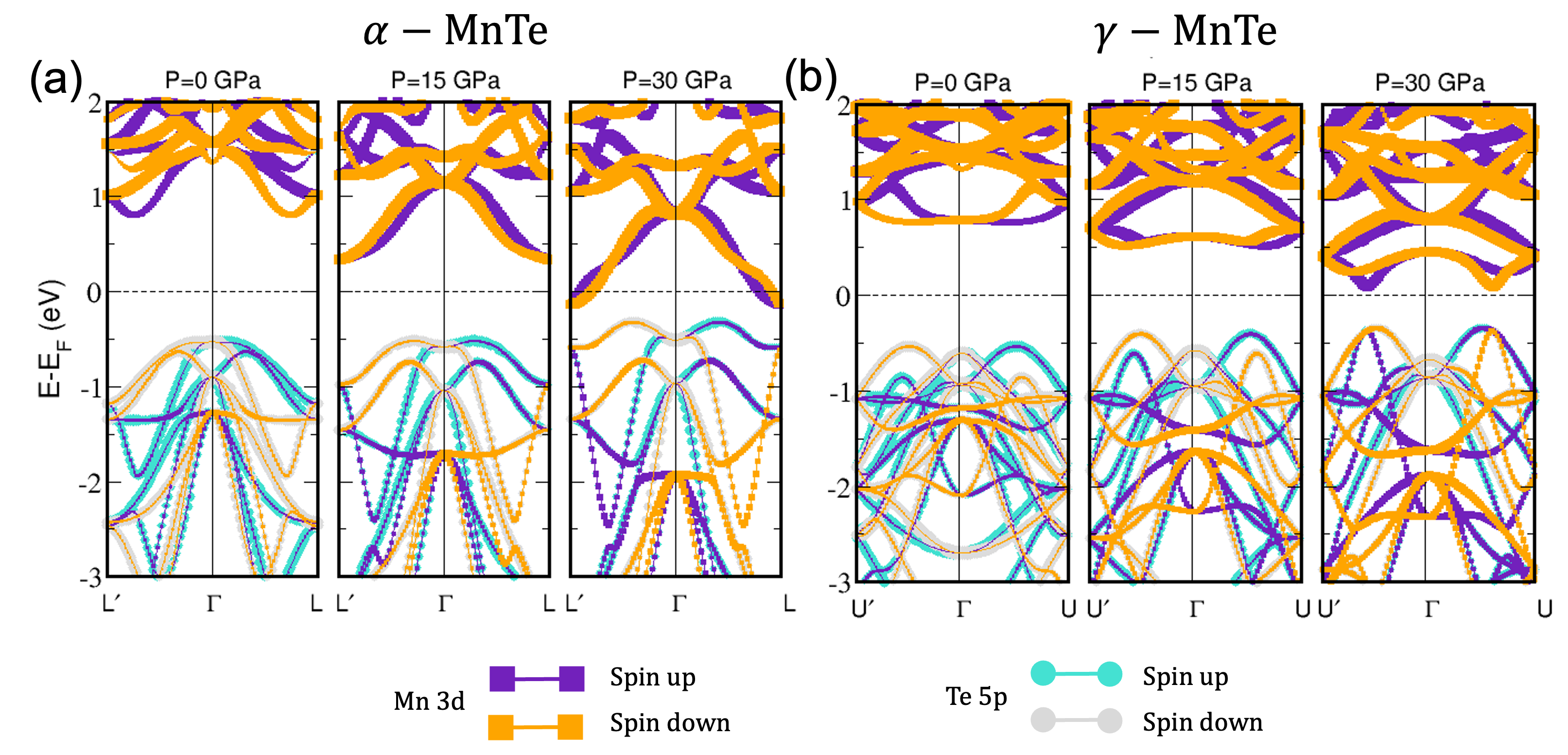}}
\caption{\textbf{Orbital-projected band structures of $\alpha$-MnTe and $\gamma$-MnTe as a function of pressure.} The orbital-projected band structure of $\alpha$-MnTe along $L'$-$\Gamma$-L (a) and $\gamma$-MnTe along the $U'$-$\Gamma$-U path (b) is depicted under various pressures. Bands projected onto spin-up and spin-down Mn 3$d$ orbitals are represented by indigo and orange squares, respectively, while those projected onto spin-up and spin-down Te 5$p$ orbitals are shown as turquoise and grey circles, respectively. In both systems the valence band arises from hybridized Te $5p$ orbitals and Mn 3$d$ orbitals, with Te 5$p$ yielding the major contribution. The contribution of Mn 3$d$ orbitals in the valence manifold increases with rising pressure. The conduction manifold is primarily attributed to Mn 3$d$ orbitals.}
\label{subfig_proj_band}
\end{figure*}

\begin{figure*}
\centerline{\includegraphics[scale=.5]{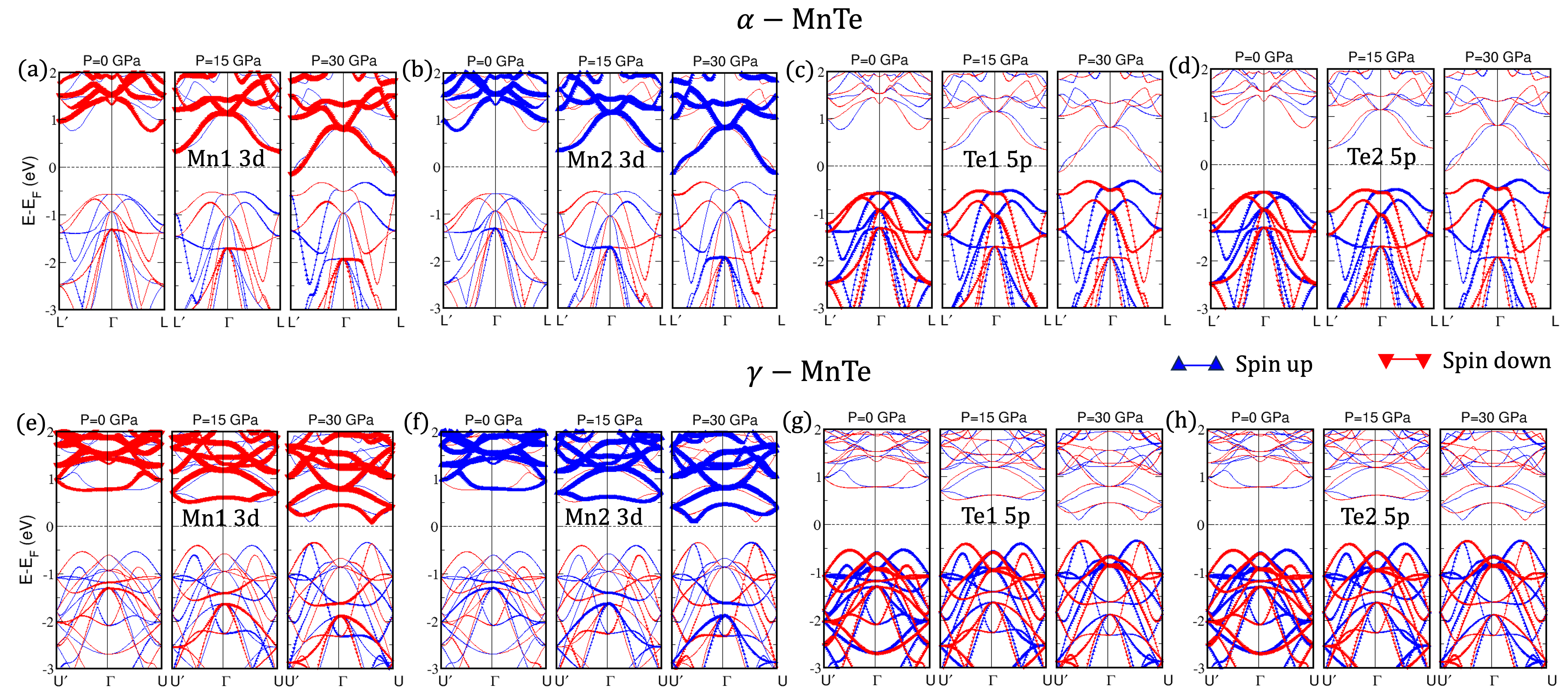}}
\caption{\textbf{Orbital-projected band structures of each atom in $\alpha$-MnTe and $\gamma$-MnTe as a function of pressure.} The orbital-projected band structures for different atoms in $\alpha$-MnTe along the $L'-\Gamma-L$ path and $\gamma$-MnTe along $U'-\Gamma-U$ under various pressures are illustrated in the top ((a)-(d)) and bottom ((e)-(h)) panels, respectively. Bands projected onto spin-up and spin-down orbitals are indicated by blue and red triangles, respectively. In both the systems, Mn1 and Mn2 atoms exhibit opposite magnetization, resulting in contrasting contributions from these atoms to the band structure. Conversely, the contributions from nonmagnetic Te1 and Te2 atoms remain consistent across the band structure.}
\label{subfig_atom_proj_band}
\end{figure*}

\begin{figure*}[t]
\centerline{\includegraphics[scale=.5]{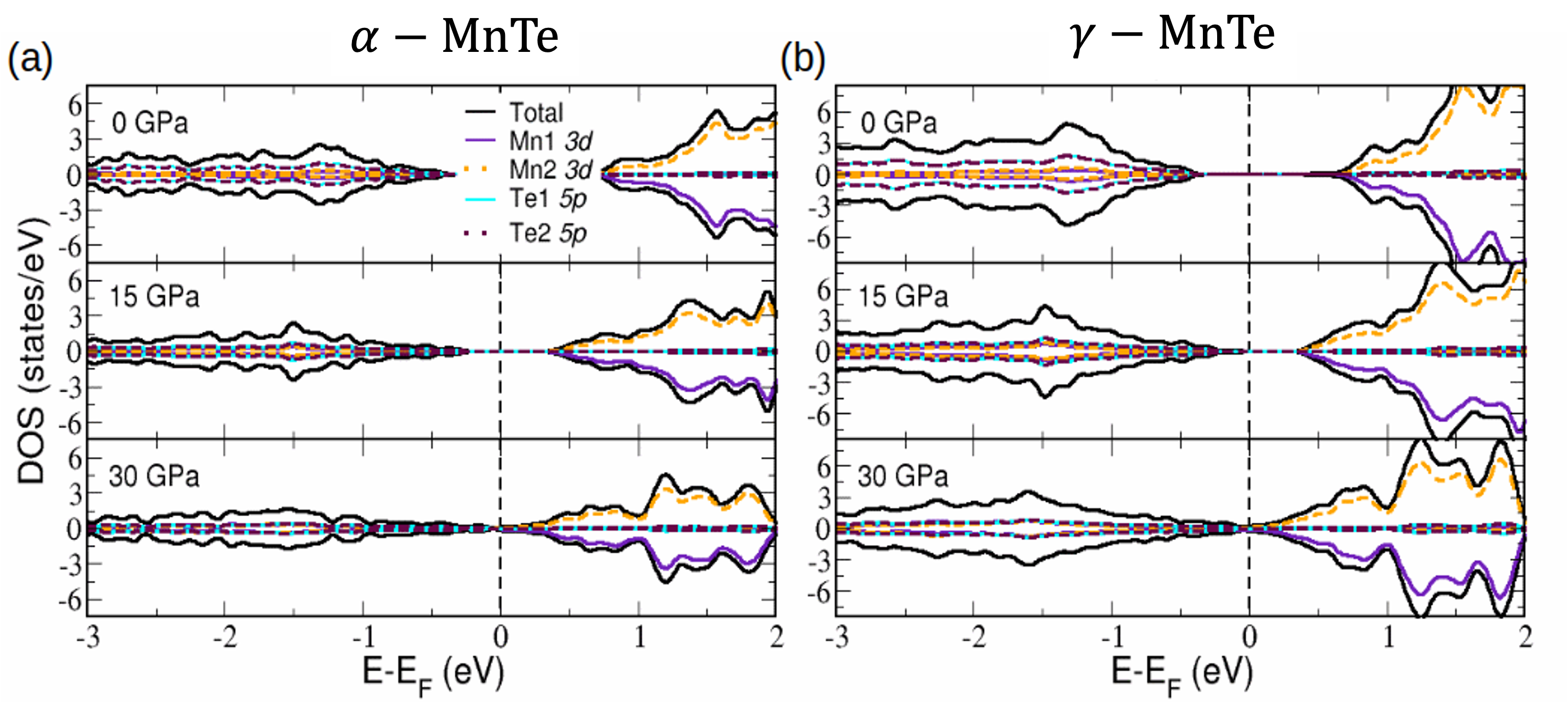}}
\caption{\textbf{Density of states of $\alpha$-MnTe and $\gamma$-MnTe as a function of pressure.} The density of states of $\alpha$-MnTe (a) and $\gamma$-MnTe (b) are illustrated at three distinct pressure values. In the valence band, both Mn1 and Mn2 contribute equally to the up-spin and down-spin channels. In contrast, in the conduction band, Mn1 contributes to the down-spin channel and Mn2 contributes to the up-spin channel. The Te 5$p$ orbital in the valence band shows greater sensitivity to pressure than Mn 3$d$ orbital. These features are similar in both MnTe structures.}
\label{subfig_PDOS_press}
\end{figure*}

\section{Details of electronic structure of $\alpha$-MnTe and $\gamma$-MnTe under pressure}
\label{SI-press-dos-band}
The atomic orbital contributions to the electronic structure of both MnTe phases are investigated as a function of pressure in order to obtain a better understanding. The projected band structures at different pressure values for $\alpha$-MnTe and $\gamma$-MnTe are shown in the (a) and (b) panels of Fig.~\ref{subfig_proj_band}. For both MnTe structures, Mn 3$d$ orbitals have an overwhelming contribution at the conduction band edges. In the valence band, at low pressures, Te 5$p$ orbital contribution is significantly greater than Mn 3$d$ orbitals. However, as pressure increases, Mn 3$d$ orbital contribution increases, indicating enhanced hybridization. This increase in hybridization can be attributed to the decrease in bond lengths between Mn and Te atoms as pressure rises [Fig.~\ref{subfig_press_parameters}(b), (f)].

The contributions of Mn1, Mn2, Te1, and Te2 to the electronic band structures (Fig.~\ref{subfig_atom_proj_band}) and DOS (Fig.~\ref{subfig_PDOS_press}) of the spin-up and spin-down channels are analyzed separately for both MnTe structures. Interestingly, in the valence band, both Mn1 and Mn2 atoms contribute equally to the spin-up and spin-down channels. From the projected band structure, Fig.~\ref{subfig_atom_proj_band} (a) and (b), it is evident that Mn1 contribution to the spin-up (down) band along $\Gamma-L(L')$ is equivalent to Mn2 contribution in the spin-down (up) band along $\Gamma-L'(L)$. However, in the conduction band, only Mn2 contributes to the spin-up band, and only Mn1 contributes to the spin-down band. Te1 and Te2 have equal states available in both spin channels across valence and conduction bands, contributing equally in each. A similar trend is observed for $\gamma$-MnTe as well [Fig.~\ref{subfig_proj_band} (e)-(f)].

Even though, orbital contributions to the valence band and conduction band, and their evolution with pressure are similar for both the $\alpha$-MnTe and $\gamma$-MnTe structures, there are significant changes observed in their band structures, including band dispersion, band crossing, band edge positions, and spin splitting. Analysing the details of projected band structure of $\alpha$-MnTe [Fig.~\ref{subfig_proj_band}(a)], it is seen that at 0 GPa, the valence band maximum is at the high symmetric point $\Gamma$, exhibiting a nearly flat nature around $\Gamma$. As pressure increases, the flatness reduces. The Mn $3d$ contribution does not increase much at this point compared to the same band at $L$. Notably, at $\Gamma$, there exists another flat band situated nearly 1.29 eV below the Fermi level, where Mn 3$d$ orbitals contribute substantially more compared to other bands. This band remains flat even at higher pressures but shifts to lower energies. For instance, at 30 GPa, it resides 1.9 eV below the Fermi level. The bands at $\Gamma$ in the conduction manifold are not flat at low pressures but become flatter at higher pressures. The energy difference between bands at $\Gamma$ increases with pressure. At low pressures, the energy differences between the valence band edge bands at $\Gamma$ and the next two bands below it are nearly 0.35 eV and 0.75 eV, respectively. At 30 GPa, these values change to 0.46 eV and 1.43 eV. In the conduction band, the energy difference between the conduction band at $\Gamma$ and the next two bands above it is 0.21 eV and 0.35 eV, respectively. At 30 GPa, these values increase to 0.5 eV and 1.3 eV. For $\gamma$-MnTe, the valence band edge remains in between high symmetric points, $\Gamma$ and $U$ at different pressure values. However, the conduction band minima at $\Gamma$ at 0 GPa changes its position to a point in between $\Gamma$ and $U$ as pressure increases. At 0 GPa, and 30 GPa, the conduction band minima is formed by different bands. Band crossing in this structure is much larger than that in the $\alpha$-MnTe.


\end{document}